\documentclass[11pt,twoside]{article}

\usepackage{asp2006}
\usepackage{epsf}
\usepackage{psfig}
\usepackage{lscape}

\begin{document}
\title{Planets in the Galactic Bulge: Results from the SWEEPS Project}   
\author{
Kailash C. Sahu,\altaffilmark{1}
Stefano Casertano,\altaffilmark{1} 
Jeff Valenti,\altaffilmark{1}
Howard E. Bond,\altaffilmark{1}
Thomas M. Brown,\altaffilmark{1}
T. Ed Smith,\altaffilmark{1}
Will Clarkson, \altaffilmark{1}
Dante Minniti,\altaffilmark{2}
Manuela Zoccali,\altaffilmark{2} 
Mario Livio,\altaffilmark{1}
Alvio Renzini,\altaffilmark{3}
R. M. Rich,\altaffilmark{4} 
Nino Panagia,\altaffilmark{1} 
Stephen Lubow,\altaffilmark{1}
Timothy Brown,\altaffilmark{5}
Nikolai Piskunov\altaffilmark{6}}

\altaffiltext{1}{Space Telescope Science Institute, 3700 San Martin Drive, 
Baltimore, MD. 21218, USA}
\altaffiltext{2}{Universidad Catolica de Chile, Av. Vicu–a Mackenna 4860,
Santiago, Chile}
\altaffiltext{3}{INAF - Osservatorio Astronomico di Padova, Vicolo 
dell'Osservatorio 5, 35122 Padova, Italy}
\altaffiltext{4}{University of California at Los Angeles, Los Angeles, CA
90095-1562, USA}
\altaffiltext{5}{Las Cumbres Observatory Global Telescope, Goleta, CA.}
\altaffiltext{6}{Department of Astronomy, Uppsala University,
Box 515, 75120 Uppsala, Sweden}

\begin{abstract}  The exoplanets discovered so far have been  mostly around
relatively nearby and bright stars. As a result, the host stars are mostly (i)
in the Galactic disk, (ii) relatively massive, and (iii) relatively metal rich.
The aim of the SWEEPS project is to extend our knowledge to stars which (i) are
in a different part of the Galaxy, (ii) have lower masses, and (iii) have a
large range of metallicities.  To achieve this goal, we used the {\it Hubble
Space Telescope\/} to monitor  180,000 F, G, K, and M dwarfs in the Galactic
bulge continuously for 7 days in order to search for transiting planets.  We
discovered 16 candidate transiting extrasolar planets with periods of 0.6 to
4.2~days, including a new class of ultra-short period planets (USPPs) with
$P<1.2$~days. Radial-velocity observations of the two brightest candidates
support their planetary nature. These results suggest that planets are as
abundant in the Galactic bulge as they are in the solar neighborhood, and they
are equally abundant around lower-mass stars (within a factor $\sim$2). The
results also suggest that planet frequency increases with metallicity even for
the stars in the Galactic bulge. All the USPP hosts are low-mass stars,
suggesting that either close-in planets around higher-mass stars are
irradiatively evaporated,or that planets are able to migrate to and survive in
close-in orbits only around such old and low-mass stars. 

\end{abstract}

\section{Introduction}   More than 250 extrasolar planets have been discovered
within the past few years, most of them through the radial velocity (RV)
measurements, and some through transits and microlensing (see J. Schneider,
Extrasolar Planet Encyclopaedia  for an up-to-date listing). These discoveries
have led to tremendous advancements in our knowledge of exoplanets. However,
the exoplanet discoveries have so far been mostly around relatively nearby and
brighter stars: all of the RV detections and a large number of transit
detections are confined to host stars within about 200 pc, a few of the transit
detections have host stars as far away as 2 pc, and the small number of the
microlensing detections have host stars as far away as 6 kpc. In addition, the
RV detections have been mostly confined to relatively higher-mass stars,
although RV studies are now being extended to M dwarfs (Marcy, 2005; Butler et
al. 2004; Bonfils et al. 2004). In contrast with the RV results, an intensive
transit search in the globular cluster 47~Tuc (Gilliland et al. 2000) found
\emph{no} hot Jupiters around $\sim$34,000 cluster members, compared to the
$\sim$17 expected from the frequency in the solar neighborhood. This
discrepancy was tentatively attributed to either environment or metallicity
effects, since 47~Tuc stars lie in a very dense stellar environment and are
significantly metal-poor compared to those in the solar neighborhood.  Indeed, 
Fischer \& Valenti (2003) find that the frequency of planets in the RV sample
rises rapidly with metallicity. So, some of the key questions in the study of
extrasolar planets, at present, are the following: (i) Are planets equally
abundant in other parts the Galaxy?  (ii) Are planets equally numerous around
lower mass stars? (iii) Are hot Jupiters common around a very different
population? (iv) Does heavy element abundance favor planet formation at other
parts of the galaxy?

Our SWEEPS (Sagittarius Window Eclipsing Extrasolar Planet Search) project was
designed to provide answers to these key questions. At a distance of $\sim 8.5$
kpc, the Galactic bulge has a large concentration of stars whose metallicities
range over $-1.5 <  [Fe/H]  <  +0.5$ (Rich and Origlia, 2005;  Zoccali et al.
2003; Fulbright et al. 2005), and hence is an ideal choice for this study.  We
used the \emph{HST} and the Wide Field Camera of the Advanced Camera for
Surveys to monitor $\sim$180,000 F, G, K and~M dwarfs with $18.5<V<26$ in a
dense stellar field ($3.3\times 3.3$ arcmin) in the Galactic bulge for transits
by orbiting Jovian-sized planets.

\begin{figure}[!ht]
\begin{center}
\plotone{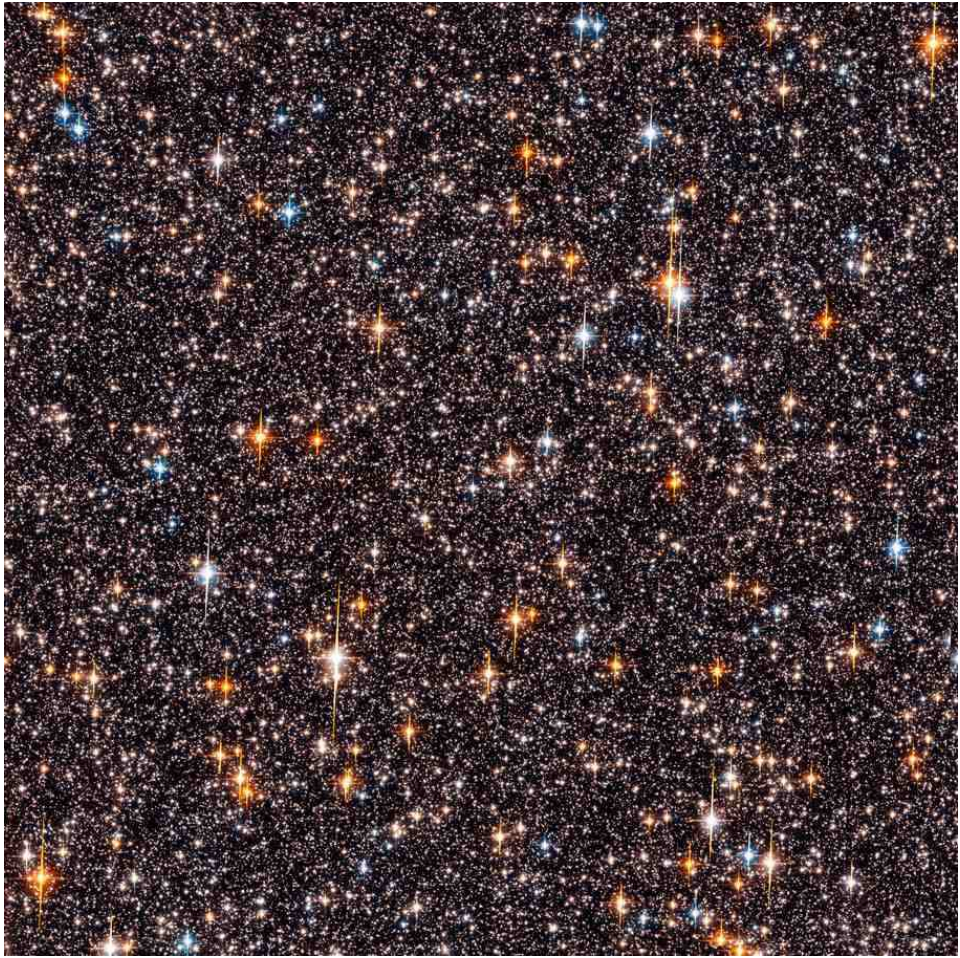}
\caption{V (F606W) and I (F814W) composite image of the SWEEPS field,
which has a size of 202 x 202 arcsec.  There are 245,000 stars down to $V\sim
30$, out
of which there are 180,000 stars brighter than $V\sim 26$ around which the
observations are sensitive to detecting Jovian planets. }
\end{center}
\end{figure}

\section{Observations}

The SWEEPS field lies in the Sagittarius-I Window of the Galactic bulge. We 
monitored this field for planetary transits over a continuous 7-day  interval
during February 22-29, 2004. At the distance of the Galactic bulge, an $M_0$
dwarf of  0.5 M$_{\odot}$ has an apparent visual magnitude of $\sim 25.5$, for
which the HST photometry is capable of detecting planetary transits.  The 
observations include 254 exposures in F606W (wide V)  and 265 exposures in
F814W (I) for the primary time series, all with an exposure time of 339 sec.

\begin{figure}[!ht]
\begin{center}
\plotone{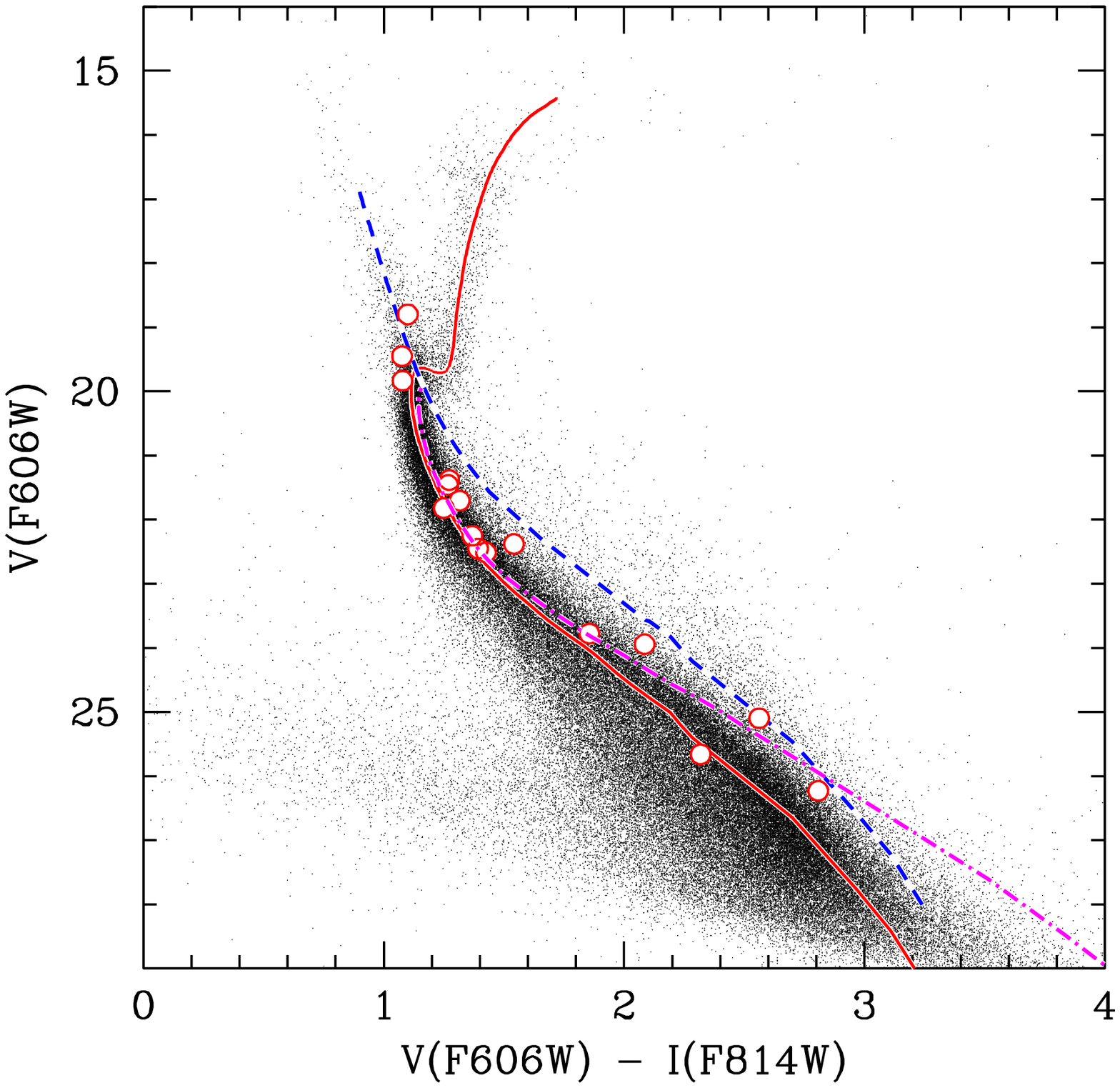}

\caption{The color-magnitude diagram (CMD) of the SWEEPS field as derived from the
deep, combined ACS images, with total integration times of 86,106 and 89,835 s in
the V and I filters, respectively.  The red (solid) line shows a 10-Gyr old
solar-metallicity isochrone which its the dominant bulge population. The  dashed
blue (upper) line shows an unevolved main sequence, representative of the
foreground young disk population. An higher-metallicity isochrone with [Fe/H]=0.5
is shown by the dashed magenta (lower) curve. Large circles represent the 16 host
stars with transiting planet candidates.} \end{center} \end{figure}

\section{Analysis}

The analysis technique employed is Difference Image Analysis (DIA; e.g., Alard
1999), similar to the procedure adapted by Gilliland et  al. (1999, 2000) for
the analysis of 47 Tuc data. Combining together all the exposures taken in each
filter using the above procedure  produces extremely deep,
twice-oversampled V (F606W) and I (F814W) images.  Figure 1 shows the combined
image of the SWEEPS field in F606W and F814W filters.

\begin{figure}[!ht] \begin{center} \plotone{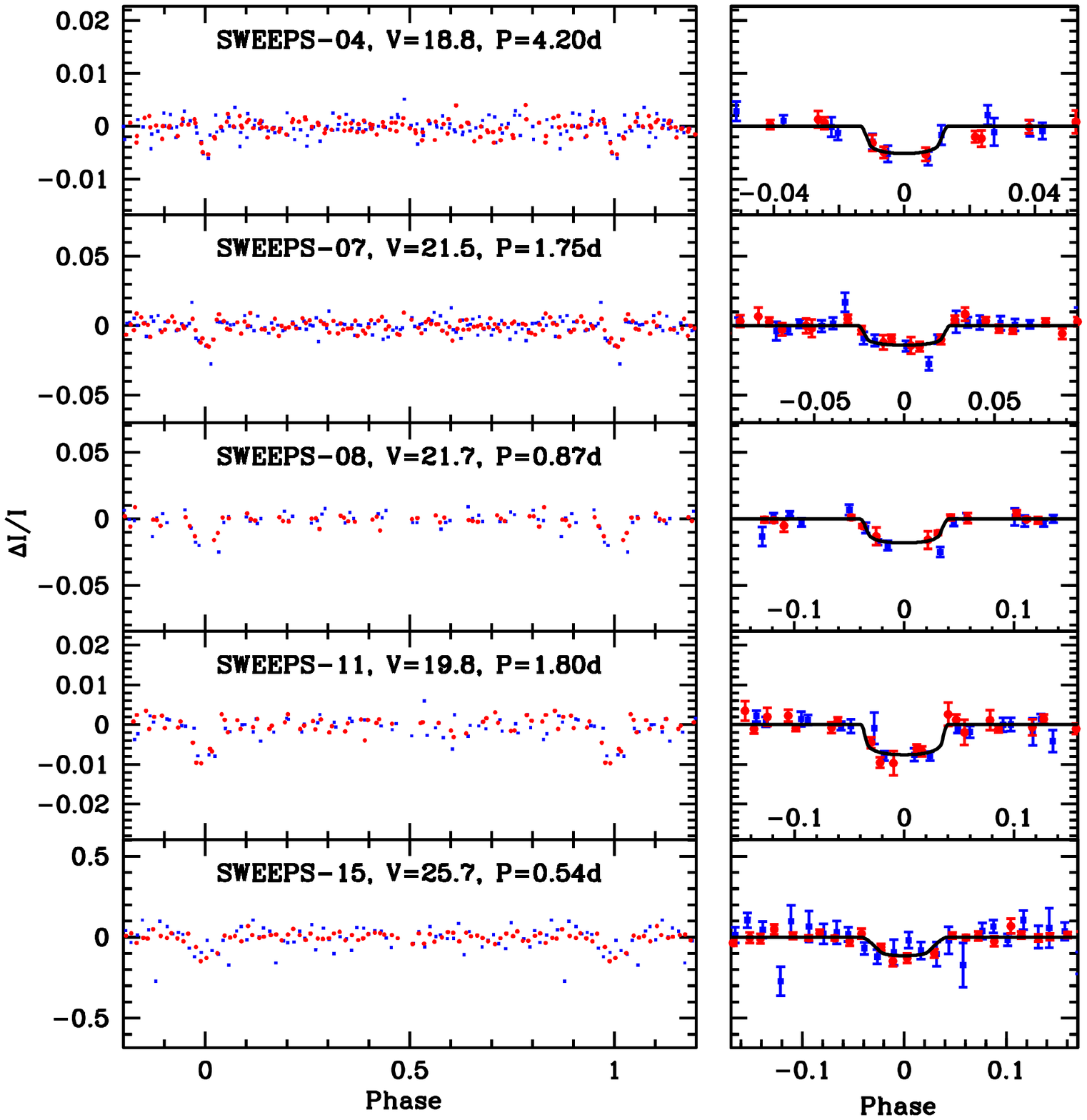} \caption{ Five
examples of observed transit light curves. The left panels show the entire
light curve, phased at the derived orbital period, and the right panels show
magnified views of the transit with 2$\sigma$ error bars. The light curves have
been binned in phase to a bin width of 1/6th of the transit duration. (Blue)
squares are the V-band observations, and (red) circles are the I-band
observations. The black solid curves are the best-fitting model transit light
curves.}  \end{center} \end{figure}

The absolute photometry (Vegamag system) of the stars in the SWEEPS field was
determined from  twice-oversampled co-added images of the entire dataset in V
and I. The DAOPHOT II PSF-fitting photometry package was used for this purpose,
with the photometric zero-points taken from the calibration work at STScI
(Sirianni et al. 2005).

About 245,000 stars are detected in this combined image down to $V \sim 30$, 
of which 180,000 stars are brighter than $V \sim 26$ around which our program
is sensitive to detecting Jovian planets.  The color-magnitude  diagram (CMD),
presented in Figure 2, shows two stellar components: a dominant  population of
old stars with a main-sequence turnoff near $V=19.6$ and well-populated
sub-giant and giant branches, and a less numerous, closer, younger and brighter
main sequence. We associate the old population with the Galactic bulge, and the
younger objects with the foreground Galactic disk (Kuijken \& Rich 2002,
Zoccali et al. 2000). A modified version of the code developed by Kovacs et al.
(2003) was used for transit search.

\section{Results and Screening for false positives}

A series of criteria as described by Sahu et al. (2006) was employed to
eliminate false positives, which include eliminating candidates with (i) a
transit depth implying a companion radius $ > 1.4 R_J$ (ii) ellipsoidal light
variations, (iii) secondary eclipses, (iv) different transit depths in V and
I. We also eliminated objects in which the photo-center of the transit signal 
is offset with respect to that of the uneclipsed star. As an additional check,
we doubled the period and re-calculated the transit depths, and eliminated
candidates with varying primary and secondary depths.  This process led to the
detection of 16 candidate planets. The magnitudes of their host stars range
from V=18.8 to 26.2, corresponding to stellar masses of 1.24 to 0.44 M$_\odot$.
Figure 3 shows a few typical examples of the observed transit light curves.  

\begin{figure}[!ht]
\begin{center}
\epsfysize=120mm \epsfbox{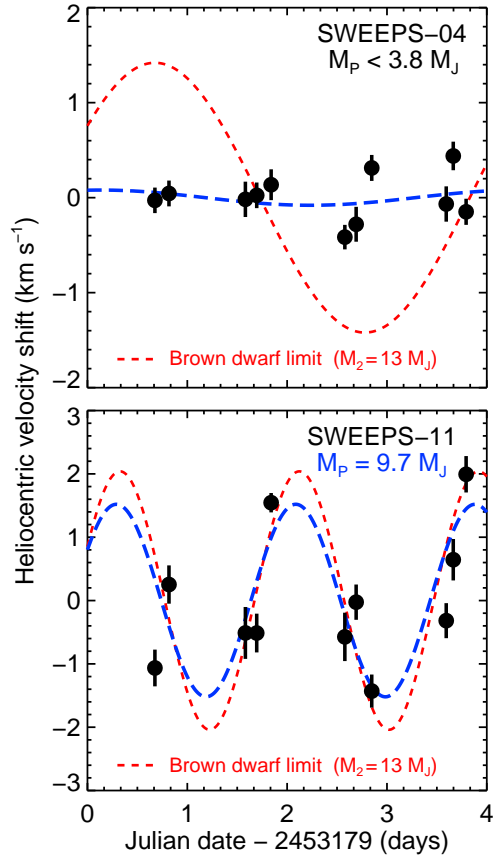}

\caption{Radial-velocity measurements of SWEEPS-04 and SWEEPS-11 from VLT
spectra.   The measured radial velocities and their associated errors are shown
as black points. The red (short-dashed) curves show the RV variation expected for a
minimum-mass brown dwarf companion of 13 M$_J$. For SWEEPS-11, there is a clear
detection of RV variations, which imply a planetary mass of  9.7 $\pm$4.5 $M_J$.
For SWEEPS-4, there is no detection, and at the 95\% and 99.9\% confidence
levels, we rule out companions more massive than 3.8 $M_J$ and 5.3 $M_J$.  The
zero-point uncertainty  in phase due to the extrapolation from the 2004 February
HST transit observations to the 2004 June date of the RV observations is 0.45
days.} \end{center} \end{figure}

In addition to the 16 exoplanet candidates, we have also detected 165 low-mass
eclipsing binaries, which we used to statistically estimate the possible
contribution from grazing eclipses and low-mass stars. Unlike most other
ground-based experiments, the HST experiment has (i) near-continuous time
coverage, (ii) observations in 2 different bands, (iii) same exposure times for
all observations, and (iv) same psf-characteristics in all the images. Such a
consistent set of observations  makes it possible statistically estimate the
contributions from astrophysical false positives (such as grazing eclipses,
low-mass stars, etc.) Furthermore, the HST observations do not suffer from
blending problems or ``red noise", which makes the detections more robust.
Taking into account all possible contributions from other sources of false
positives, we estimate that $\ga 45\%$ of the candidates are genuine
planets (See Sahu et al, 2006 and 2008 for more details).

\begin{figure}[!ht]
\begin{center}
\plotone{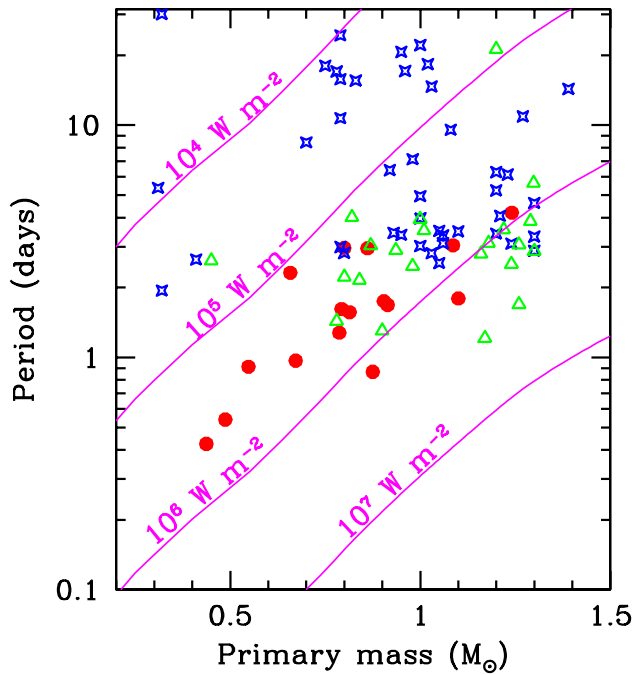}

\caption{Orbital periods and host-star masses for extrasolar planets with
periods up to $\sim$12 days. Solid (red) circles are the 16  SWEEPS candidates, (green)
triangles are transiting planets around brighter stars as derived from
ground-based observations, and (red) crosses are for planets detected through RV
variability. The SWEEPS candidates extend the range of planetary orbital periods
down to 0.42 days. Very few planets have irradiances above  $2 \times 10^6 W
m^{-2}$ which corresponds to an equilibrium temperature of 2000 K.  None in the
SWEEPS sample have equilibrium temperatures larger than 2000 K.  The absence of
ultra-short-period planets around stars $>0.9 M_\odot$ may be due to irradiative
evaporation. }

\end{center}
\end{figure}

\section{Radial Velocity Followup Observations} Most of the host stars are too
faint for radial velocity followup observations, but SEEPS-4 and SWEEPS-11 were
bright enough and lie in a relative uncrowded region so that we could obtain
radial velocity observations of them, using the ESO 8m VLT and the FLAMES/UVES
spectrograph. For SWEEPS-11, we clearly detected RV variations, which indicate
the mass to be 9.7 $M_J$. For SWEEPS-4, for which the transit detection has a
high S/N, the RV variations were below the detection limit suggesting an upper
limit to its mass of 3.8 $M_J$.  If only 50\% of our candidates are genuine
planets, the probability that both selected objects would be planets is 25\%.
If 30\% of the candidates are genuine planets, this probability is only 10\%.
This gives us extra confidence that a large fraction must be planets, and
supports our estimate that $\ga 45\%$ of the candidates are genuine
planets. .

\section{Results}

After correcting for geometric transit probability and our detection
efficiency, our detections suggest that the frequency of planets in the SWEEPS
field is similar to that in the local neighborhood.

The frequency of planets around low-mass stars is also similar to the frequency
of planets around higher-mass stars, but given the small number statistics, the
uncertainty is large which can easily be a factor of 2 or 3.

The host stars of the detected planets preferentially lie towards
higher-metallicity isochrones. This is consistent with the fact that metallicity
favors planet frequency in the Galactic bulge, similar to the findings in the
solar neighborhood. 

The USPPs with orbital periods shorter than 1 day occur only around stars less
massive than 0.88 M$_\odot$, and which have preferentially higher-metallicity.
This suggests that planets orbiting very close to more massive stars might be
evaporatively destroyed, or that  planets can migrate to close-in orbits and
survive there only around such old and low-mass stars.

\acknowledgements We would like to thank Ron Gilliland for his help on several
stages of the project.

\end{document}